\documentclass[11pt,a4paper]{article}
\usepackage{imports}
\usepackage{generalmacros}
\newcommand{\bterr}{\mathrm{err_{BT}}}

\newcommand{\Lap}{\mathrm{Lap}}

\newcommand{\eps}{\ensuremath{\epsilon}}
\newcommand{\epsinv}{\ensuremath{\eps^{-1}}}

\newcommand{\apxd}{\ensuremath{\widetilde{d}}}

\usepackage{bbm}

\begin{document}
\date{}

\title{Tighter Bounds for Local Differentially Private Core Decomposition and Densest Subgraph}

\author[1]{Monika Henzinger}
\author[2]{A. R. Sricharan}
\author[3]{Leqi Zhu}
\affil[1,3]{Institute of Science and Technology, Austria}
\affil[2]{Faculty of Computer Science, UniVie Doctoral School of Computer Science DoCS, University of Vienna, Austria}
\maketitle
\date{}
\begin{abstract}
Computing the core decomposition of a graph is a fundamental problem that has recently been studied in the differentially private setting, motivated by practical applications in data mining. In particular, \cite{DLR22kcore} gave the first mechanism for approximate core decomposition in the challenging and practically relevant setting of local differential privacy.
One of the main open problems left by their work is whether the accuracy, i.e., the approximation ratio and additive error, of their mechanism can be improved.
We show the first lower bounds on the additive error of approximate and exact core decomposition mechanisms in the centralized and local model of differential privacy, respectively.
We also give mechanisms for exact and approximate core decomposition in the local model, with almost matching additive error bounds. Our mechanisms are based on a black-box application of continual counting.
They also yield improved mechanisms for the approximate densest subgraph problem in the local model. \end{abstract}

\section{Introduction}
\label{sec:intro}

Core decomposition and densest subgraph are fundamental problems that can be used to identify structure within graphs, with wide-ranging practical applications (see~\cite{DLR22kcore}). Recently, motivated by scenarios where the graph encodes confidential user information, there has been a growing interest in developing ``privacy-preserving'' algorithms for these problems. In particular, research has focused on algorithms that satisfy \emph{differential privacy}~\cite{dwork2006calibrating}, the de facto standard notion of privacy, which requires the distribution of the algorithm's output to be nearly identical on graphs that differ by a single edge. Intuitively, this ensures that its output reveals little about the edges of the underlying graph.

Such algorithms have been designed for both the \emph{centralized} and \emph{local} models of differential privacy. In the centralized model, a \emph{trusted} party runs the algorithm, with full access to all user data (here, the entire graph). The local model removes the trust assumption and is more distributed in nature. In this model, each user (here, a vertex in the graph) interactively discloses information about its data (here, its incident edges) in multiple \emph{rounds} of communication with an \emph{untrusted} server\footnote{The server may accidentally leak their messages, e.g., due to security breaches. However, it is not malicious.}, which produces the output. All messages sent throughout the interaction, not just the output, must be differentially private. 

In both models, a primary metric of interest is the \emph{accuracy} of the output. This is fairly well understood in the centralized model, where there are fast and private (approximation) algorithms for core decomposition~\cite{dhulipala2023near} and densest subgraph~\cite{nguyen2021differentially,FHS21dense,DLR22kcore,dhulipala2023near,dinitz2023improved} with optimal and near-optimal additive error, respectively. However, there are still large gaps in our understanding of this metric (and its trade-off with round complexity) in the local model. In a groundbreaking paper, \cite{DLR22kcore} initiated the study of both problems in the local model by showing various trade-offs between approximation ratio, additive error, and round complexity. In particular, they gave (1) a private $(2+\eta)$-approximate algorithm for core decomposition that has $O(\log^3 n)$ additive error and $O(\log n)$ round complexity, as well as (2) a private $(4+\eta)$-approximate algorithm for densest subgraph with the same trade-off. 
More recently, \cite{dinitz2023improved} gave private exact and $(2+\eta)$-approximate algorithms for densest subgraph with improved additive error, roughly $O(\log^2 n)$, and with $O(n^2\log n)$ and $O(\log n)$ round complexity, respectively (see also Table~\ref{tab:results}). 
All aforementioned error bounds hold \emph{with high probability}, i.e., with probability at least $1-n^{-c}$, for any fixed constant $c > 0$.

It is not clear, however, whether these are the best possible trade-offs.
Thus, we seek to understand the following question: 
\begin{quote}
\emph{What is the minimum additive error achievable, with high probability, for differentially private core decomposition (and densest subgraph) on $n$-vertex graphs, in the local model}?
\end{quote}

\subsection{Our contributions}

\subparagraph*{Lower bounds.} We prove a tight lower bound on the additive error of approximate core decomposition in the centralized model, which carries over to the (less powerful) local model, regardless of round complexity.  We also take the first step towards proving an accuracy and round complexity trade-off for exact core decomposition, by showing that 1-round algorithms must be inaccurate. To the best of our knowledge, these are the first lower bounds for core decomposition in either model. 
Specifically, we show the following results (see also Table~\ref{tab:results}). For simplicity, we consider mechanisms satisfying pure differential privacy, but similar bounds should hold for approximate differential privacy.

\begin{itemize}
\item \emph{Lower bound for centralized and local model}
(Theorem~\ref{thm:coreness-centralized-lb}). For constant $\gamma \geq 1$, we show that any \emph{centralized} algorithm for $\gamma$-approximate core decomposition has $\Omega(\gamma^{-1}\log n)$  additive error, with constant probability. This is tight for exact algorithms~\cite{dhulipala2023near}. Since centralized algorithms can simulate local ones, the bound also holds in the local model, regardless of the round complexity of the algorithm. 

\item \emph{Lower bound for 1-round local model}
(Theorem~\ref{thm:non-interactive-coreness-lb}). We show that any local algorithm for exact core decomposition that uses a single round of communication has $\Omega(\sqrt{n})$ additive error, with constant probability, on a large family of graphs.
\end{itemize}

\subparagraph*{Upper bounds.}
To show that our first lower bound (which is already tight in the centralized model) is unlikely to be improved in the local model, we give local algorithms for exact and $(2+\eta)$-approximate core decomposition with improved, resp.~nearly matching error,
which exhibit new trade-offs with round complexity. In particular, we show that it is possible to achieve an additive error that does not have a linear dependence on the round complexity, answering a question of~\cite{DLR22kcore}. 

Using a known (approximate) reduction from densest subgraph to core decomposition, our results for core decomposition immediately lead to local algorithms for $2$-approximate and $(4+\eta)$-approximate densest subgraph with the same trade-offs. 

The main idea of our mechanisms is that the users each independently  employ a \emph{continual counting} mechanism designed for the centralized model. Note that improving the accuracy of such a mechanism (to match the known lower bound for continual counting~\cite{DNPR10dpco}) would immediately improve the accuracy of our algorithms to match our lower bound, i.e., to optimal. 

Finally, we point out a trade-off between memoryless algorithms and non-memoryless ones. 
In {\em memoryless local} algorithms~\cite{DLR22kcore}, the users do not have persistent memory (between rounds), apart from their private information. 
This might be motivated by the local model used 
for differentially private hypothesis testing and learning, which employs interactive protocols between the users and the curator where no user remembers its computation in prior rounds when computing the output for the current round.
We show that certain counting mechanisms can be emulated memorylessly. Others appear to require memory (i.e., each user needs to remember values from prior rounds), but can produce better error bounds in local algorithms with memory, depending on the maximum degree $\Delta$ of the graph. Showing a separation of the two models is an interesting open problem. 
Specifically, we show the following results (see also Table~\ref{tab:results}). We note that all our mechanisms satisfy pure differential privacy.

\begin{itemize}
\item {\em Exact core decomposition}
(Theorem~\ref{thm:peeling}). We give the first memoryless local algorithm for exact core decomposition that has $O(\log^2 n)$ additive error, with high probability, and $O(n)$ round complexity. With memory, the error improves to $O(\log n \log \Delta)$, with high probability, which is nearly optimal when $\Delta = O(\mathrm{polylog}(n))$.
\item {\em $(2+\eta)$-approximate core decomposition}
(Theorem~\ref{thm:apx-peeling}). For constant $\eta > 0$, we give a memoryless local algorithm for $(2+\eta)$-approximate core decomposition that has nearly optimal additive error, $O(\log n \log\log n)$, with high probability, and $O(\log^2 n)$ round complexity.
\item {\em Approximate densest subgraph}
(Theorem~\ref{thm:densest-subgraph}). For constant $\eta > 0$, we give memoryless local algorithms for $2$-approximate and $(4+\eta)$-approximate densest subgraph\footnote{Here, the algorithm returns a set of vertices, which induce a subgraph whose density is approximately maximal.} with $O(\log^2 n)$ and $O(\log n \log \log n)$ error, with high probability, and round complexity $O(n)$ and $O(\log^2 n)$, respectively. With memory, the error of the $2$-approximate algorithm improves to $O(\log n \log \Delta)$, with high probability. 
\end{itemize} 

\begin{table}[t]
\footnotesize
\centering
\begin{tabular}{cccccc}
\toprule
\textbf{Problem} & \textbf{Apx. Factor} & \textbf{Additive Error}  & \textbf{Rounds} & \textbf{Memoryless?} & \textbf{Citation}  \\
\midrule
\multirow{4}{*}{\textbf{Core Decomposition}} & $1$ & $O(\log^2 n)$ & $O(n)$ & \cmark & [this work]  \\
& $1$ & $O(\log n\log \Delta)$ & $O(n)$ & \xmark & [this work]  \\
& $1$ & $\Omega(\sqrt{n})$ & 1 & \xmark & [this work]  \\
& $2+\eta$ & $O(\log^3 n)$ & $O(\log n)$ & \cmark & \cite{DLR22kcore}  \\
& $2+\eta$ & 
$O(\log n\log \log n)$ & $O(\log^2 n)$ & \cmark & [this work] \\
& $\gamma$ & $\Omega(\gamma^{-1}\log n)$ & any & \xmark & [this work] \\ \midrule
\multirow{6}{*}{\textbf{Densest Subgraph}} & $1$ & $O(\log (n\delta^{-1})\log n)$ & $O(n^2\log n)$ & \cmark & \cite{dinitz2023improved}  \\
& $2$ & $O(\log^2 n)$ & $O(n)$ & \cmark & [this work]  \\
& $2$ & $O(\log n\log \Delta)$ & $O(n)$ & \xmark & [this work]  \\
& $2+\eta$ & $O(\eta^{-1}\log^2 n)$ & $O(\eta^{-1}\log n)$ & \cmark & \cite{dinitz2023improved}  \\
& $4+\eta$ & $O(\log^3 n)$ & $O(\log n)$ & \cmark & \cite{DLR22kcore}  \\
& $4+\eta$ & $O(\log n \log\log n)$ & $O(\log^2 n)$ & \cmark & [this work]  \\
\bottomrule
\end{tabular}
\caption{Summary of error bounds in the local model. Each upper bound is for a mechanism satisfying $(\epsilon,\delta)$-edge differential privacy and holds with high probability. If $\delta$ appears in the expression, then $0 < \delta < n^{-\Omega(1)}$; otherwise, $\delta = 0$. Lower bounds are for $\epsilon$-differential privacy and hold with constant probability. To save space, the dependency on $\epsilon$ is omitted from each additive error upper bound.}
\label{tab:results}
\end{table}
\subsection{Technical overview}

\subparagraph*{Centralized lower bound.} Our lower bound for $\gamma$-approximate core decomposition in the centralized model uses a standard ``packing argument''
\cite{HT10dpgeometry}. The idea is to construct a collection of disjoint ``bad'' output classes for a certain input and show that the algorithm has a ``large enough'' probability of producing an output in each of the bad classes, due to differential privacy. The only subtlety is that we need to carefully construct the output classes to take advantage of the approximation guarantee. 

\subparagraph*{Local lower bound.} Our lower bound for 1-round, exact core decomposition in the local model is obtained by reduction to a problem in the \emph{centralized} model, for which there is a strong lower bound. Specifically, it is known that, to privately answer $\Theta(n)$ \emph{random} inner product queries on a secret $\{0,1\}^n$ vector, most responses need to have $\Omega(\sqrt{n})$ additive error~\cite{de2012lower}. Recently, \cite{DBLP:conf/icalp/EdenLRS23} gave an elegant lower bound on the additive error of 1-round triangle counting algorithms in the local model using similar techniques. 

In our case, we define a class of ``query graphs'', one per inner product query, in which the coreness of a \emph{fixed} vertex $x$ (i.e., its assigned ``score'' in the core decomposition) is (roughly) the answer to the query. 
In addition to $x$, there are some \emph{secret} vertices (and other vertices). For each secret vertex $v$, the existence of the edge $\{v,x\}$ is private information (which depends on the secret vector). 
Crucially, all neighborhoods of $x$ and the secret vertices in {\em any possible query graph}  %
appear in two specific query graphs, namely, ones corresponding to the all-ones and all-zeros query vectors, respectively. The neighborhoods of the remaining vertices do not depend on the secret vector. 

Our approach to solving the inner product problem is now as follows. The centralized algorithm first simulates the 1-round local core decomposition algorithm on the two fixed graphs to determine the messages that $x$ and the secret vertices would send in {\em any} query graph and saves these messages.
Subsequently, when answering a query, the centralized algorithm \emph{reuses} the saved messages for $x$ and the secret vertices (without further privacy loss) when it simulates the core decomposition algorithm on the corresponding query graph. This allows it to answer many queries  correctly. 

Originally, we had a more complex, direct argument. We briefly mention it here to give an idea of what is going on ``under the hood'' of the reduction. Specifically, we showed that, on a class of random graphs (similar to the query graphs instantiated in our reduction), most transcripts of a 1-round core decomposition algorithm have the property that, conditioned on seeing the transcript, the coreness of some (fixed) vertex still has high variance, $\Omega(n)$. This implies that the standard deviation of the additive error is $\Omega(\sqrt{n})$. We prefer the reduction argument, as it is simpler and gives a stronger result.

\subparagraph*{Local upper bounds.} On the algorithmic side, our starting point is the $(2+\eta)$-approximate core decomposition algorithm of~\cite{DLR22kcore}, which essentially implements an approximate version of the classic, \emph{exact} peeling algorithm for core decomposition. To remove the approximation factor, it is natural to consider implementing the original algorithm, which is what we do to obtain our local algorithm for exact core decomposition. 
At a high level, the main technical challenge here is for each user (vertex) to privately disclose its degree as the server continually deletes subsets of users from the graph (based on the disclosed degrees). In~\cite{DLR22kcore}, each user simply adds fresh Laplace noise to its actual degree and discloses the resulting sum. In this case, the standard deviation of the noise (which dominates the additive error) needs to be {\em linear in  the number of disclosures}, i.e., deletions, to ensure privacy. This is acceptable for the approximate peeling algorithm, where $O(\log^2 n)$ deletions occur, but not for the exact algorithm, where $\Omega(n)$ deletions might occur. 

A \emph{continual counting} mechanism~\cite{DNPR10dpco} privately discloses all (noisy) prefix sums of an input sequence of integers. There are \emph{adaptive} variants, where the elements of the input sequence arrive online, one at a time, and the mechanism discloses the sum after each arrival. Our new approach is for each user to use an adaptive continual counting mechanism to track the  \emph{sum of its degree changes} due to  deletions. It can then disclose its degree as its (noisy) initial degree plus the last sum. 

There are several points that make implementing this non-trivial. (a) First, note that each user runs its own continual counting mechanism whose input (apart from its neighborhood list) is given by the server. Thus, if the inputs to the mechanisms (on two graphs that differ by a single edge) can differ for each user, even slightly, then the total privacy loss can be $\Omega(n)$, which would require each mechanism to add prohibitively large noise to maintain overall privacy. Through a careful analysis, we show that the total privacy loss is constant. (b) Next, since the outputs of the counting mechanisms have noise, the users do not report their actual degrees, but rather \emph{noisy} ones, so the error could amplify over time. We prove that the exact peeling algorithm is \emph{robust}, in the sense that the error of its output is bounded by the maximum error of the noisy degrees (which is fairly small, with high probability).
(c)
Lastly, in the definition of the local model given by~\cite{DLR22kcore}, users do not have persistent memory. We show that the \emph{binary-tree} counting mechanism can be emulated without persistent memory. On the other hand, we do not know how to do this for the \emph{sparse-vector} based counting mechanism, which gives better parameterized error bounds, namely, in terms of the maximum degree $\Delta$ of the graph. 

Our local algorithm for $(2+\eta)$-approximate core decomposition is obtained by applying the same thought process to the $(2+\eta)$-approximate algorithm of~\cite{DLR22kcore}. The difference here is that, since the number of deletions is $O(\log^2 n)$, the counting mechanisms have to disclose far fewer prefix sums and, hence, achieve better accuracy. 

Finally, \cite{dhulipala2023near} showed that their exact core decomposition can be post-processed to obtain a 2-approximate densest subgraph with roughly the same additive error (in the centralized model). Similarly, \cite{DLR22kcore} implicitly proved this for their $(2+\eta)$-approximate core decomposition, obtaining a $(4+\eta)$-approximate densest subgraph (in the local model). We generalize their arguments to any approximation ratio, filling in details of the proof of~\cite{dhulipala2023near} (see the paragraph before Lemma~\ref{lem:density-coreness-accuracy}). Specifically, we give a sufficient condition that allows a $\gamma$-approximate core decomposition to yield a $2\gamma$-approximate densest subgraph with the same (asymptotic) additive error (in either model). Then we show that the condition is satisfied by our $1$-approximate and $(2+\eta)$-approximate algorithms, hence obtaining $2$-approximate $(4+\eta)$-approximate densest subgraph algorithms, respectively, with the same trade-offs.

\subparagraph*{Related work.} 
There is a centralized algorithm for exact core decomposition with additive error $O(\log n)$~\cite{dhulipala2023near}. Similar to our local algorithm, this algorithm privately implements the classic, exact peeling algorithm. However, it uses different techniques (a multi-dimensional version of the sparse vector technique), which seem difficult to translate to the local model.
It is known that any exact densest subgraph algorithm in the centralized model has $\Omega(\sqrt{\log n})$ additive error~\cite{nguyen2021differentially,FHS21dense}. To the best of our knowledge, only certain private exact core decomposition algorithms can be post-processed to yield 2-approximate densest subgraph algorithms with the same accuracy (see Lemma~\ref{lem:density-coreness-accuracy}). Hence, the above lower bound does not apply to all exact core decomposition algorithms. Continual counting has been used in the centralized model, along with other techniques, to obtain a $(2+\eta)$-approximate private densest subgraph algorithm with additive error $O(\log^{2.5}n)$, with high probability, which runs in near-linear time~\cite{FHS21dense}. There, the privacy analysis has a similar flavor as ours, but is more complex, while the accuracy analysis is quite different. Finally, there are insertions-only and deletions-only graph algorithms under continual observation that have used the idea of tracking the difference sequence of a desired quantity (e.g.~degree of a vertex) via continual counting in the centralized model~\cite{DBLP:conf/esa/FichtenbergerHO21}.

\section{Preliminaries}
\label{sec:prelim}

\subparagraph*{Core decomposition and densest subgraph}

In this paper, a \emph{graph} is always simple (i.e., no self-loops or parallel edges), undirected, and unweighted. Let $G = (V,E)$ be a graph. We use the notation $V(G)$ and $E(G)$ to denote the set of vertices and edges of $G$, respectively. A graph $H$ is a \emph{subgraph} of $G$ if $V(H) \subseteq V(G)$. Given a subset $U$ of vertices of $G$, the subgraph of $G$ \emph{induced} by $U$, denoted $G[U]$, is the \emph{largest} subgraph $H$ of $G$ such that $V(H) = U$, i.e., $|E(H)|$ is maximized. A vertex $u$ of $G$ is \emph{adjacent} to (or, a \emph{neighbor} of) a vertex $v \neq u$ of $G$ if and only if $\{u,v\} \in E(G)$. The \emph{degree} of a vertex $v$ in $G$, denoted $\deg_G(v)$, is the number of vertices adjacent to $v$ in $G$. Two graphs $G$ and $G'$ (on the same set of vertices) \emph{differ} on a single edge (say $e$) if the symmetric difference of their edge sets, $(E(G) \setminus E(G')) \cup (E(G') \setminus E(G))$, consists of a single element (namely, $e$).

The \emph{coreness} of a vertex $v$ in a graph $G$, denoted $k_G(v)$, is the largest integer $k$ such that there exists a subgraph $H$ of $G$ such that, for every vertex $u \in V(H)$, the (induced) degree of $u$ in $H$ is at least $k$, i.e., $\deg_H(u) \geq k$. We omit the subscript when it is clear from the context. For integer $k \geq 0$, the \emph{$k$-core} of the graph $G$ is the subgraph of $G$ induced by the vertices of $G$ with coreness at least $k$ in $G$, i.e., $G[U]$, where $U = \{ v \in V \mid k_G(v) \geq k \}$\footnote{An alternative, equivalent definition is that the $k$-core of $G$ is the (unique) largest subgraph of $G$ in which every vertex has degree at least $k$.}. In the \emph{core decomposition problem}, the input is a graph $G = (V,E)$ and the goal is to return a vector $(k_G(v) : v \in V)$ containing the coreness of each vertex of $G$ (assuming a fixed ordering of the vertices of $G$).

The \emph{density} of a graph $G$, denoted $\rho(G)$, is the quantity $|E(G)|/|V(G)|$. A \emph{densest subgraph} of $G$ is a subgraph $H$ of $G$ such that $\rho(H)$ is maximized, among all subgraphs of $G$. Notice that, for any subset  $U$ of vertices of $G$, if $H$ is a subgraph of $G$ such that $V(H) = U$, then $\rho(H) \leq \rho(G[U])$. In the \emph{densest subgraph} problem, the input is a graph $G = (V,E)$ and the goal is to return a subset of vertices $U \subseteq V(G)$ that induce a densest subgraph of $G$.

\subparagraph*{Differential privacy} For any $\epsilon,\delta > 0$, a randomized algorithm $\mathcal{A}$ mapping (inputs in) $\mathcal{X}$ to (outputs in) $\mathcal{Y}$ is \emph{$(\epsilon,\delta)$-differentially private}~\cite{dwork2006calibrating} if for all \emph{neighboring} inputs $x,x' \in \mathcal{X}$ and for all subsets $Y \subseteq \mathcal{Y}$ of outputs, $\Pr(\mathcal{A}(x) \in Y) \leq e^\epsilon \Pr(\mathcal{A}(x') \in Y) + \delta$.  Here, the definition of neighboring depends on the context and the problem. If $\delta=0$, we simply write \emph{$\epsilon$-differentially private}.

\begin{fact}
Let $\mathcal{A}$ be an $\epsilon$-differentially private algorithm mapping $\mathcal{X}$ to $\mathcal{Y}$. 
\begin{enumerate}
    \item (\emph{Composition}) Suppose that $\mathcal{B}$ is an  $\epsilon'$-differentially private algorithm mapping $\mathcal{X} \times \mathcal{Y}$ to $\mathcal{Z}$. Then the \emph{composition} of $\mathcal{A}$ and $\mathcal{B}$, i.e., the randomized algorithm mapping $\mathcal{X}$ to $\mathcal{Z}$ via $x \mapsto \mathcal{B}(x,\mathcal{A}(x))$, is $(\epsilon+\epsilon')$-differentially private.
    \item (\emph{Post-processing}) Suppose that $\mathcal{B}$ is a randomized algorithm mapping $\mathcal{Y}$ to $\mathcal{Z}$. Then the \emph{post-processing} of $\mathcal{A}$ by $\mathcal{B}$, i.e., the randomized algorithm mapping $\mathcal{X}$ to $\mathcal{Z}$ via $x \mapsto \mathcal{B}(\mathcal{A}(x))$, is $\epsilon$-differentially private.
\end{enumerate}  
\end{fact}

The \emph{$\ell_1$-sensitivity} of a function $f : \mathcal{X} \to \mathbb{R}^d$, denoted $\Delta_1(f)$, is the supremum of the quantity $\|f(x) - f(x')\|_1 = \sum_{i=1}^d |f(x)_i - f(x')_i|$, over all neighboring $x,x' \in \mathcal{X}$. The \emph{Laplace} distribution (centered at 0) with \emph{scale} $b > 0$ has probability density function $f(x) = \frac{1}{2b} \exp(-|x|/b)$. We write $X \sim \mathsf{Lap}(b)$ or just $\mathsf{Lap}(b)$ to denote a random variable $X$ that is distributed according to the Laplace distribution with scale $b$.

\begin{fact}
\label{lem:laplace_tailbounds}
If random variable $X \sim \mathsf{Lap}(b)$, then, for any $\beta > 0$, $\Pr[|X| > b \log (1/\beta)] \le \beta$.
\end{fact}

\begin{fact}[Theorem 3.6 in \cite{journals/fttcs/DworkR14}] \label{lem:Laplacemech} Let $f : \mathcal{X} \to \mathbb{R}^d$ be any non-constant function and let $\epsilon > 0$. Suppose that, for each $i \in [d]$, random variable $X_i \sim \Lap(\Delta_1(f)/\epsilon)$. Then the \emph{Laplace mechanism} 
$
\mathcal{A}(x)=f(x)+(X_1,\dots,X_d)
$
is $\epsilon$-differentially private.
\end{fact}

\subparagraph*{Local differential privacy}

In the \emph{local} model of differential privacy, there are $n$ \emph{users}, each with \emph{private data}, who communicate in synchronous \emph{rounds} with a \emph{server} (or \emph{curator}). In each round, each user can send a message to the server, based on its private data, its local memory, the messages (from the server) that it has received so far, and possibly some \emph{fresh} local randomness. Upon receiving all user messages, the server can either broadcast a message to all users, or end the interaction and produce an output. If it broadcasts a message, then the users will receive it at the end of the round. Subsequently, they may write information to their local memories, which persists into future rounds, before continuing to the next round. Similarly, the server also has local memory.

A \emph{local mechanism} $\mathcal{M}$ specifies (randomized) algorithms for the server and for each user, which dictate the messages they send in each round and what they write to their local memories; $\mathcal{M}$ is \emph{memoryless} if each user never writes to its local memory. The \emph{input} of $\mathcal{M}$ is the private data of each user. An \emph{execution} of $\mathcal{M}$ on an input generates a \emph{transcript}, which consists of the messages sent by the server and all users in every round, if they run the algorithms specified by $\mathcal{M}$, with each user having the corresponding private input data. The \emph{round complexity} of a mechanism is the maximum number of rounds in any transcript. A local mechanism is \emph{$\epsilon$-differentially private} if it is $\epsilon$-differentially private, when viewed as a randomized algorithm mapping private user data to transcripts (provided the notion of neighboring user data is defined).

When the local model is applied to graphs, there is an underlying \emph{input graph}, $G$, whose vertices are the users (and public knowledge). The private data of each user (vertex) is its set of neighbors in $G$. Hence, only users $u$ and $v$ know whether $\{u,v\}$ is an edge in $G$. Two input graphs are \emph{neighboring} if they differ in at most one edge. In other words, a local mechanism in this setting is $\epsilon$-\emph{(edge) differentially private} if its transcript is $\epsilon$-differentially private on neighboring input graphs.

The prevailing formalization of local differential privacy used for graph data~\cite{DLR22kcore,DBLP:conf/icalp/EdenLRS23,dinitz2023improved} was adapted from the general (non-graph) setting~\cite{ledp}. 
In this formalism, \emph{local randomizers} are used to describe the computations carried out by the users to generate the (differentially private) messages they send in each round. Specifically, an \emph{$\epsilon$-local randomizer} is a randomized algorithm $R$ that takes as input a subset of vertices of the input graph and returns an output such that, for all subsets of vertices $N,N'$ (corresponding to possible neighborhoods of a user) that differ in at most one element and any subset $Y$ of outputs, $\Pr[R(N) \in Y] \leq e^{\epsilon} \Pr[R(N') \in Y]$. 
Paraphrasing~\cite{DLR22kcore}, a \emph{local mechanism} is specified by a potentially infinite set of local randomizers, $\mathcal{R}$, together with a function, $\mathcal{A}$, describing how the interaction proceeds. A \emph{$0$-round transcript} is the empty sequence. Given a $t$-round transcript $\pi$, $\mathcal{A}(\pi)$ either returns $\bot$, indicating the interaction ends, or a pair $(S_U^{t+1},S_R^{t+1},S_\epsilon^{t+1})$ encoding the set $S_U^{t+1}$ of users who participate in round $t+1$, the set $S_R^{t+1} \subseteq \mathcal{R}$ containing the local randomizer assigned to each participating user, and the corresponding privacy parameters $S_\epsilon^{t+1}$. 
Suppose that $S_O^{t+1}$ contains the outputs generated by the local randomizers after each participating user runs its assigned local randomizer on its private data (i.e., neighborhood in the input graph) in round $t+1$. Then the concatenation $\pi \odot (S_U^{t+1},S_R^{t+1}, S_\epsilon^{t+1}, S_O^{t+1})$ is a \emph{$(t+1)$-round} transcript. In Appendix~\ref{sec:prelim-extra}, we give a similar formalism for our model (reflecting the description above) and show that memoryless local mechanisms in our model can be emulated using local randomizers. However, it seems difficult to emulate persistent memory as some random choices stored in user local memory cannot be made public\footnote{The \emph{sparse vector technique} (SVT), for one, is an important mechanism that requires random choices to be stored in local memory. The noise added to the threshold $T$ in SVT requires adding a single noise upfront to $T$ (giving $\tilde T$), and this $\tilde T$ is maintained until the threshold is crossed. Importantly, $\tilde T$ \emph{cannot} be disclosed. Using fresh random noise for the threshold in each iteration increases the error to linear in the number of iterations, which is fatal.}. We describe our local mechanisms in terms of server and user algorithms, instead of local randomizers.

\subparagraph*{Continual counting}
\label{sec:countingprelims}

In the \emph{continual counting} problem, the input is a sequence of integers (a \emph{stream}) up to some fixed, finite length, $T$. The goal is to output all prefix sums of the stream. A \emph{continual counting} mechanism allows the elements of the input stream to be \emph{inserted}, one at a time. After each insertion, the mechanism outputs the \emph{count} (prefix sum) so far. Two streams (of the same length) are \emph{neighboring} if they differ in at most one element, by at most one. The definition of differential privacy for neighboring input streams captures non-adaptive counting mechanisms. We refer to~\cite{JRSS23pricedpco} for a formal definition of \emph{adaptive} counting mechanisms, where the elements of the stream can depend on the outputs of the mechanism.

The \emph{binary-tree} mechanism is a differentially private adaptive continual counting mechanism that was introduced by \cite{DNPR10dpco} and \cite{CSS10bintreemech}. Using the sparse-vector technique, \cite{DNRR15dprect} gave an improved version of the binary tree mechanism, which we call the \emph{sparse-vector} counting mechanism, where the error at a time step is a function of the true count at that time step.

\begin{restatable}{theorem}{continualcounting}[\citealp{DNPR10dpco, CSS10bintreemech}]
\label{thm:continual-counting}
Let $\epsilon > 0$ be a constant. There is an $\epsilon$-differentially private adaptive counting mechanism such that, with probability at least $1 - \beta$, at any single time step $t \in [T]$, the error is bounded by
\[
\bterr(t, \beta, \epsilon) = O \left( \epsilon^{-1} \cdot \log t  \sqrt{\log 1/\beta} \cdot \max \left\{ \sqrt{\log t}, \sqrt{\log 1/\beta} \right\} \right)
\]
\end{restatable}

\begin{restatable}{theorem}{sparsecontinualcounting}[\citealp{DNRR15dprect}]
\label{thm:sparse-continual-counting}
Let $\epsilon > 0$ be a constant. There is an $\epsilon$-differentially private adaptive counting mechanism such that, with probability at least $1 - \beta$, at any single time step $t \in [T]$, the error is bounded by
\[
O \left(
\bterr( \min \{ n_t, t \}, \beta, \epsilon)
+
\epsilon^{-1}\cdot \log (t/\beta)
\right)
\] where $n_t$ is the true continual count at time $t$ and $\bterr$ is the error of the binary tree mechanism.
\end{restatable}

\section{Core decomposition via continual counting}
\label{sec:peelalg}

Let $\epsilon > 0$ be any positive constant. In this section, we describe an $\epsilon$-edge differentially private mechanism for core decomposition on $n$-vertex graphs, in the local model. The mechanism runs for at most $n$ rounds in any execution and has $O(\epsilon^{-1}\log^2 n)$ additive error with high probability. It can also be made memoryless. In the following, we refer to the coreness of a vertex that is returned by the mechanism as the \emph{estimated} coreness of that vertex.

\begin{restatable}{theorem}{peeling}
\label{thm:peeling}
For any $\epsilon > 0$, there is a memoryless local $\epsilon$-edge differentially private mechanism $\mathcal{M}$ that returns an \emph{estimate} of the coreness of each vertex in an $n$-vertex graph $G = (V,E)$ such that:
\begin{itemize}
    \item $\mathcal{M}$ runs for at most $n$ rounds in any execution, and
    \item for each vertex $v \in V$ with actual coreness $k(v)$, the estimated coreness $\tilde{k}(v)$ of $v$ satisfies for $\alpha = O(\epsilon^{-1} \log^2 n)$ with high probability:
    \[
        k(v) - \alpha \leq \tilde{k}(v) \leq k(v) + \alpha.
    \]
\end{itemize}
With memory, $\alpha = O(\epsilon^{-1}\log n \log \Delta)$ with high probability, where $\Delta$ is the maximum degree of $G$.
\end{restatable}

At a high level, we implement a private version of the classic \emph{peeling algorithm} of Matula and Beck~\cite{matula1983smallest}. This algorithm deletes vertices in increasing order of their coreness by repeatedly performing the following steps, until the graph is empty:
\begin{enumerate}
    \item compute the minimum degree, $d$, among all vertices currently in the graph,
    \item iteratively delete vertices with degree at most $d$ from the graph, until none are left, and
    \item report the coreness of each vertex deleted in the preceding step as $d$.
\end{enumerate}

To do so, a natural approach is for each vertex to send its \emph{noisy} degree to the server in each round. That is, it sends its degree, plus some random noise to ensure privacy. Given these noisy degrees, the server can determine the set of vertices to be deleted in that round and broadcast it to every vertex. Upon receiving the server's message, each (surviving) vertex can determine which of its neighbors have been deleted and update its degree accordingly.

The main difficulty with this approach is generating the noisy degrees. In particular, a vertex may participate in many rounds (e.g.,~$\Omega(n)$ on a line graph) and, hence, send many noisy degrees. Na\"{i}vely using fresh noise each time requires too much noise to account for the privacy loss.

Our observation is that substantially less noise is needed if each vertex $v$ uses a private \emph{continual counting} mechanism, $\mathcal{C}_v$, instead. Specifically, $v$ first generates a noisy degree by adding Laplace noise to its initial degree and sends this in the first round. If $v$ is not deleted in a round, then it inserts the (absolute) \emph{change} in its degree, as a result of zero or more of its neighbors being deleted in the round, into $\mathcal{C}_v$. Its initial noisy degree, minus the count that it receives from $\mathcal{C}_v$, is the noisy degree that it sends in the next round.

Algorithms~\ref{alg:peeling} and~\ref{alg:peelingv} contain the pseudocode for the server and vertices, respectively. For simplicity, we assume that each vertex $v$ maintains the state of $\mathcal{C}_v$ using persistent memory; we discuss how to remove this assumption at the end.

\begin{algorithm}[!ht]
\SetAlgoLined
\DontPrintSemicolon
\caption{Server algorithm for coreness estimation on a graph $G = (V,E)$, in the local model.}
\label{alg:peeling}
$d \gets 0$\;
\ForEach{round $t=1,2,\dots,n$}{
    \textbf{receive} noisy degree $\tilde{d}_t(v)$ from each vertex $v \in A_t := V \setminus (S_1 \cup \dots \cup S_{t-1})$\;
    $d \gets \max \{ d, \min \{ \tilde{d}_t(v) \mid v \in A_t\} \}$\;
    $S_t \gets \{ v \in A_t \mid \tilde{d}_t(v) \leq d \}$\;
    \textbf{broadcast} $S_t$ to all vertices in $A_t$ and set estimate $\tilde{k}(v) = d$ for each vertex $v \in S_t$\;
    \lIf{$V \setminus (S_1 \cup \dots \cup S_t) = \emptyset$}
        {\textbf{return} estimate vector $(\tilde{k}(v) : v \in V)$}
}
\end{algorithm}

\begin{algorithm}[!ht]
\SetAlgoLined
\DontPrintSemicolon
\caption{User algorithm for coreness estimation on a graph $G = (V,E)$ in the local model; code for each vertex $v \in V$. We assume that $v$ is given its set of neighbors, $N_v$, in $G$, as input and that $v$ maintains an $\epsilon/2$-differentially private continual counting mechanism, $\mathcal{C}_v$, supporting up to $n$ adaptive insertions.}
\label{alg:peelingv}
$\tilde{d}_1(v) \gets |N_v| + \mathsf{Lap}(4/\eps)$\;
\ForEach{round $t = 1,2, \dots,n$} {
    \textbf{send} $\tilde{d}_t(v)$ to server and \textbf{receive} message $S_t$ from server\;
    \lIf{$v \in S_t$} {\textbf{terminate}}
    $\tilde{d}_{t+1}(v) \gets \tilde{d}_1(v) - \mathcal{C}_v.\textsc{Insert}(|N_v \cap S_t|)$\;
}
\end{algorithm}

\subsection{Privacy}

The key point is as follows: if the transcript (containing the sequence of outputs of each continual counting mechanism) is the same for the first $t-1$ rounds in two executions of the mechanism on neighboring graphs, then the sequence of inputs inserted into each continual counting mechanism is nearly the same in both executions.

\begin{observation}
\label{obs:core-sens}
Suppose the mechanism is executed on two graphs that differ by an edge $e$. If the transcript is the same in the first $t-1$ rounds in both executions, then the following holds.
\begin{enumerate}
    \item The set of vertices deleted (i.e., the message broadcast by the server) in each round $r < t$ is the same in both executions.
    \item For each vertex that is not an endpoint of $e$, its initial degree and subsequent changes in degree are the same in each round $r < t$ in both executions.
    \item For each endpoint $v$ of $e$, its initial degree differs by one in the two executions. It has a change in degree that differs (by one) in a round $r < t$ only if the other endpoint was deleted in round $r-1$, and $v$ itself was not deleted in round $r-1$ (or earlier).
\end{enumerate}
\end{observation}

Since the initial noisy degree is differentially private, the transcript of the first round is differentially private. By the preceding observation, to ensure that a transcript of subsequent rounds has roughly equal probability of occurring when the mechanism is executed on two neighboring graphs, it suffices for the counting mechanisms to have roughly equal probability of producing the same outputs, when executed on streams that \emph{collectively} differ in at most one input (namely,
at most one change in degree of an endpoint of the differing edge). A subtle point is that each counting mechanism needs to support \emph{adaptive} insertions, as each input to the mechanism depends on its previous outputs. The next lemma formally proves this.

\begin{restatable}{lemma}{peelprivacy}
\label{lem:peelprivacy}
If the continual counting mechanism of each vertex is $\epsilon/2$-differentially private and supports up to $n$ adaptive insertions, then the transcript is $\epsilon$-edge differentially private.
\end{restatable}
\begin{proof}
We view a transcript as a vector $\pi = (\pi_1,\dots,\pi_n)$ of functions $\pi_t : V \to \mathbb{R} \cup \{\bot\}$, for $t \in [n]$, where $\pi_t(v)$ is the message sent by vertex $v \in V$ in round $t$, or $\bot$ if it sends no message. In particular, since the message, $S_t$, broadcast by the server in round $t$ is a deterministic function of $(\pi_1,\dots,\pi_t)$, we omit it from the transcript.

Consider two graphs $G$ and $G'$ that differ in an edge $e$. Let $\Pi$ and $\Pi'$ denote the (random) transcript of the mechanism when executed on $G$ and $G'$, respectively. We claim that for any set of (valid) transcripts $\mathcal{A}$, $\Pr(\Pi \in \mathcal{A}) \leq e^{\epsilon}\Pr(\Pi' \in \mathcal{A})$. Indeed, fix a transcript $\pi \in \mathcal{A}$. We will show that $\Pr(\Pi = \pi) \leq e^{\epsilon} \Pr(\Pi'  = \pi )$.
We split the analysis into two steps, each consisting of one mechanism that we analyze separately:
(1) The mechanism ${\cal M}_1$ that discloses $\tilde{d}_1(v)$ for each vertex $v$, and (2) the  mechanism  ${\cal M}_2$ that discloses $\tilde{d}_t(v)$ for each vertex $v$ and all $t>1$.

(1) Consider a vertex $v \in V$. If $v \not\in e$, then the degree of $v$ is the same in both graphs, while for the two endpoints of $e$, the difference is exactly 1. Consider the vector containing the degrees of all vertices, in some fixed order. Then the $\ell_1$-norm of the difference of the vector for $G$ and the vector for $G'$ is 2. Thus, adding Laplacian noise with parameter $\epsilon/4$ to each entry in the vector guarantees that  ${\cal M}_1$ is
$\eps/2$-edge differentially private.

(2) 
For each vertex $v \in V$, let
\begin{align*}
p_v &:= \prod_{t=2}^n \Pr(\Pi_t(v) = \pi_t(v) \mid \wedge_{r=1}^{t-1} \Pi_r  = \pi_r) \text{ and }  
p_v' := \prod_{t=2}^n \Pr(\Pi'_t(v) = \pi_t(v) \mid \wedge_{r=1}^{t-1} \Pi'_r  = \pi_r) \, .
\end{align*}

By definition, $\Pi_1(v)$ is the value returned by ${\cal M}_1$. For $t > 1$, conditioned on the transcript of the first $t-1$ rounds being $(\pi_1,\dots,\pi_{t-1})$ in both executions, the sequence of messages, $S_1,\dots,S_{t-1}$, broadcast by the server in the first $t-1$ rounds is the same in both executions.
If $v$ participates in round $t$, i.e., $v \notin S_1 \cup \dots \cup S_{t-1}$, then $\Pi_t(v)$ is the count returned by $\mathcal{C}_v$ immediately after $v$ inserts the input $c_t(v) := -|N_v \cap S_{t-1}|$ (its $(t-1)$th change in degree in the execution on $G$). Otherwise, $\Pi_t(v) = \bot$.

It follows that $p_v$ is the probability that $\mathcal{C}_v$ returns the sequence of outputs $(\pi_1(v),\pi_2(v),\dots)$ on the (adaptive) sequence of inputs $(c_1(v),c_2(v),\dots)$. Defining $c_1'(v),c_2'(v),\dots$ similarly for the execution on $G'$, we have that $p_v'$ is the probability that $\mathcal{C}_v$ returns the sequence of outputs $(\pi_1(v),\pi_2(v),\dots)$ on the sequence of inputs $(c_1'(v),c_2'(v),\dots)$.

By Observation~\ref{obs:core-sens}, the inputs to $\mathcal{C}_v$ are the same in both executions conditioned on the previous outputs being the same, i.e.,$(c_1(v),c_2(v),\dots) = (c'_1(v),c'_2(v),\dots)$ for each vertex $v \notin e$. By coupling the random bits of each vertex $v \notin e$ in the two executions, this implies that each output of $\mathcal{C}_v$ has the same probability of occurrence under both executions (conditioned on the previous outputs being equal). It follows that $p_v = p_v'$ for $v \notin e$ and, hence,
$
\prod_{v \notin e} p_v = \prod_{v \notin e} p_v' \, .
$

On the other hand, for each endpoint $v \in e$,  
and $t > 1$, $c_t(v) \neq c'_t(v)$ only if $S_{t-1}$ contains the other endpoint $u \in e$, but not $v$. In this case, 
$(c_2(v),c_3(v), \dots)$ and $(c_2'(v),c_3'(v),\dots)$ differ by one element, namely $|c_t(v)-c_t'(v)| = 1$) and $(c_2(u),c_3(u), \dots)$ equals $(c_2'(u),c_3'(u),\dots)$.
If this is case does not happen, i.e., if $u$ and $v$ belong to the same set $S_{t-1}$, then $(c_2(v),c_3(v), \dots)$ and $(c_2'(v),c_3'(v),\dots)$ do not differ.
Since $\mathcal{C}_u$ and $\mathcal{C}_v$ are $\epsilon/2$-differentially private, it follows that $p_u = p_u'$ and $p_v \leq e^{\epsilon/2}p_v'$. Hence,
$
\prod_{v \in e} p_v \leq e^{\epsilon/2}\prod_{v \in e} p'_v \, .
$

Finally, since each vertex independently generates its message in each round $t$:
\begin{align*}
\Pr(\Pi = \pi)
&= \prod_{v \in V} \prod_{t=1}^n  \Pr(\Pi_t(v) = \pi_t(v) \mid \wedge_{r=1}^{t-1} \Pi_r = \pi_r) 
= \prod_{v \notin e} p_v \times \prod_{v \in e} p_v
\leq \prod_{v \notin e}p'_v \times e^{\epsilon/2}\prod_{v \in e} p'_v \\
&= e^{\epsilon/2} \prod_{v \in V} \prod_{t=1}^n  \Pr(\Pi'_t(v) = \pi_t(v) \mid \wedge_{r=1}^{t-1} \Pi'_r = \pi_r) = e^{\epsilon/2} \Pr(\Pi' = \pi). 
\end{align*}
This shows that ${\cal M}_2$ is $\eps/2$-edge differentially private.
 By simple composition, the complete mechanism, which consists of the output of ${\cal M}_1$ and ${\cal M}_2$, is $\eps$-edge differentially private.
\end{proof}

\subsection{Accuracy} The noisy degrees sent to the server may differ significantly from the actual degrees. Hence, in each round, the server may incorrectly delete some vertices or fail to delete others. We show that if there is a bound on the error of the noisy degrees (of all vertices and in all rounds), then the same bound holds on the error of the coreness estimates.

\begin{restatable}{lemma}{peelsmooth}
\label{lem:peelsmooth}
Let $G=(V,E)$ be a graph.
If each noisy degree differs from its corresponding actual degree by at most $\alpha$, then, for every vertex $v \in V$ with actual coreness $k(v)$, the estimated coreness $\tilde{k}(v)$ of $v$ satisfies 
$k(v) - \alpha \le \tilde{k}(v) \le k(v) + \alpha$.
Furthermore, every vertex in $G[U]$ has (induced) degree at least $\tilde{k}(v) - \alpha$, where $U$ is the set of all vertices $u$ with estimated coreness $\tilde{k}(u) \geq \tilde{k}(v)$. 
\end{restatable}
\begin{proof}
Fix any vertex $v \in V$. Let $t$ be the round in which $v$ is deleted, let $d_t$ (resp.~$\tilde d_t(v)$) be the value of the variable $d$ (resp.~message from vertex $v$) stored by (resp.~received by) the server at the end of round $t$, and let $r$ be the first round in which the server sets $d$ to $d_t$. By definition, $\tilde{k}(v) = d_t$, $r \leq t$, and $d_t$ is the minimum noisy degree sent in round $r$, i.e., $\tilde{d}_r(u)\ge d_t$ for each $u \in U$, where $U$ is the set of non-deleted vertices at the start of round $r$. Since $r$ is the first round where $d$ is set to $d_t$ and $d$ is non-decreasing, it follows that $U$ is the set of all vertices $u$ with estimated coreness $\tilde{k}(u) \geq d_t$. By definition, $v \in U$  and every vertex in $G[U]$ has (\emph{actual}) induced degree at least $d_t - \alpha$ since
$\tilde{d}_r(u) \ge d_t$ and the noisy degree differs in every round at most $\alpha$ from its corresponding degree. 
By definition, $G$ contains a subgraph $H$ containing $v$ in which every vertex has induced degree at least $k(v)$. 
Note that $k(v)$ is the largest integer for which this holds, so $k(v) \geq d_t - \alpha$ and, hence, $\tilde{k}(v) =d_t \leq k(v) + \alpha$. 
This shows the second inequality of the lemma.

Consider the first vertex, $u$, of $H$ to be deleted. Let $t'$ be the round in which $u$ is deleted and let $d_{t'}$ be the value of $d$ stored at the end of round $t'$. By definition, no vertex in $H$ (in particular, $v$) has been deleted at the start of round $t'$. Hence, $t' \leq t$ and $u$ sends a noisy degree $\tilde{d}_{t'}(u)$ that is at least $k(v) - \alpha$. On the other hand, since $u$ is deleted in round $t'$, the noisy degree that it sends in that round is at most $d_{t'}$. Since $d$ is non-decreasing, it follows that $k(v)-\alpha \leq d_{t'} \leq d_t = \tilde{k}(v)$. This shows the first inequality of the lemma.
\end{proof}
Note that the second statement of Lemma~\ref{lem:peelsmooth} is only needed for the accuracy proof of the approximate densest subgraph algorithm in \cref{sec:dense}.
By Laplace tail bounds with $\beta = 1/n^{\Omega(1)}$  (Fact~\ref{lem:laplace_tailbounds}) and a union bound over all vertices, with high probability, $\deg_G(v) - O(\epsinv\log n) \le \apxd_1(v) \le \deg_G(v) + O(\epsinv\log n)$ for all $v \in V$. Hence, the additive error of $\apxd_{t+1}(v)$ is $O(\epsinv\log n)$, plus the error of the continual counting mechanism, with high probability. Note that the length of any input sequence is bounded by $T = n$ and its prefix sums are bounded by $n_T = \Delta$. Hence, taking $\beta = n^{-\Omega(1)}$, the binary tree and sparse-vector counting mechanism have additive error $O(\eps^{-1} \log^2 n)$ and $O(\eps^{-1} \log n \log \Delta )$, respectively, with high probability (Theorem~\ref{thm:continual-counting} and Theorem~\ref{thm:sparse-continual-counting}). Combining this with Lemma~\ref{lem:peelsmooth} immediately yields the following.

\begin{lemma}
\label{lem:peelaccuracy}
With high probability, for every vertex $v \in V$, the estimated coreness of $v$ differs from the actual coreness of $v$ by at most 
\begin{enumerate}
    \item $O(\eps^{-1} \log^2 n)$ when using the binary tree counting mechanism, and
    \item $O(\eps^{-1} \log n \log \Delta )$ when using the sparse-vector counting mechanism.
\end{enumerate}
\end{lemma}

\subsection{Memoryless modification}
\label{sec:memoryless_modif}
We show how to make the binary tree mechanism for each vertex memoryless.
Suppose that in round $t$, the server broadcasts the \emph{entire} transcript of messages thus far, i.e., it sends the sequence of messages, $\tilde{d}_1(v),\dots,\tilde{d}_{t-1}(v)$, sent by each vertex $v$ and the sets of vertices, $S_1,\dots,S_{t-1}$, deleted in each round. Then, using this information, each vertex $v$ can determine the sequence of inputs that it previously inserted into $\mathcal{C}_v$  (namely, $|N_v|, |N_v \cap S_1|, \dots, |N_v \cap S_{t-2}|$) and the sequence of responses it received from $\mathcal{C}_v$ (namely, $\tilde{d}_1(v),\dots,\tilde{d}_{t-1}(v)$).

Note that the state of the binary tree continual counting mechanism (i.e., the values stored at each node in the binary tree) after processing $t-1$ inputs is fully determined by the sequence of inputs to the mechanism and the respective outputs of the mechanism. 
Hence, by using this mechanism, each vertex $v$ can determine the state of $\mathcal{C}_v$ from the server's message, without having to explicitly store it. Given this, it can simulate $\mathcal{C}_v$ on the input $|N_v \cap S_{t-1}|$ to determine the message, $\tilde{d}_t(v)$, to send in round $t$. We give more details in Appendix~\ref{sec:memory_simulation}.

In the language of $\eps$-local randomizers, our result shows that each vertex can simulate the binary tree mechanism using local randomizers by returning a \emph{partial binary tree} to the server in each round, and reconstructing the binary tree in the next round using the past transcript which is encoded, for instance, in the choice of local randomizer that it has to use in that round.

\begin{lemma}
    If each vertex uses the binary tree continual counting mechanism, then the coreness estimation mechanism can be made memoryless.
\end{lemma}

\subsection{Further applications}

\cite{DLR22kcore} gave a memoryless local $\epsilon$-edge differentially private mechanism by building on \cite{LSY22kcore} for approximate core decomposition. We show in \cref{sec:apx-core} that our framework can be used in this setting to obtain the following guarantee.

\begin{restatable}{theorem}{apxpeeling}
\label{thm:apx-peeling}
For any $\epsilon, \eta > 0$, there is a memoryless local $\epsilon$-edge differentially private mechanism $\mathcal{M}$ that returns an estimate of the coreness of each vertex in an $n$-vertex graph $G = (V,E)$ such that:
\begin{itemize}
    \item $\mathcal{M}$ runs for $O(\log_{1+\eta} n\log n)$ rounds in any execution, and
    \item for each vertex $v \in V$ with actual coreness $k(v)$, the estimated coreness $\tilde{k}(v)$ of $v$ satisfies:
\[
    k(v) - \alpha \leq \tilde{k}(v) \leq (2+\eta)k(v) + \alpha \, ,
\]
where $\alpha = O(\epsilon^{-1}\log n\log \log_{1+\eta} n)$ with probability $1-n^{-\Omega(1)}$.
\end{itemize}

\end{restatable}

We also show in \cref{sec:dense} how to obtain the following guarantees for densest subgraph using our results for coreness approximation.

\begin{restatable}{theorem}{densestsubgraph}
    \label{thm:densest-subgraph}
    For any $\epsilon, \eta > 0$, there are memoryless local $\epsilon$-edge differentially private mechanisms $\mathcal{M}$ and $\mathcal{M}'$ that return subsets of vertices $U$ and $U'$ from a given $n$-vertex graph $G$, respectively, such that:
    \begin{itemize}
        \item $\mathcal{M}$ and $\mathcal{M}'$ run for at most $n$ and $O(\log_{1+\eta} n \log\log n)$ rounds, respectively, in any execution.
        \item With probability $1-n^{-\Omega(1)}$, the density of $G[U]$ and $G[U']$ is at least $\rho^*/2 - \alpha$ and $\rho^*/(4+\eta) - \alpha'$, respectively, where $\rho^*$ is the maximum density of any subgraph of $G$, $\alpha = O(\epsilon^{-1}\log^2 n)$, and $\alpha' = O(\epsilon^{-1}\log_{1+\eta} n\log \log n)$.
    \end{itemize}
    With memory, $\alpha = O(\epsilon^{-1}\log n\log \Delta)$ with high probability, where $\Delta$ is the maximum degree of $G$.
\end{restatable} %
\section{Lower bounds for core decomposition}
\label{sec:non-interactive}

Let $\epsilon > 0$ be any positive constant. We first give a lower bound in the centralized model (that carries over to the local model) and then give a lower bound for the local model.

\label{app:packinglb}

\begin{theorem}
\label{thm:coreness-centralized-lb}
Let $\gamma \ge 1$ be a constant and let $V$ be a set of $n$ vertices. Suppose that $\mathcal{M}$ is an $\epsilon$-edge differentially private mechanism in the centralized model that estimates the coreness of every vertex in a given graph $G$ on $V$ such that, for all vertices $v \in V$ with actual coreness $k(v)$, the estimated coreness $\tilde{k}(v)$ of $v$ satisfies
\[
\gamma^{-1} k(v) - \alpha \le \tilde{k}(v) \le \gamma k(v) + \alpha  \text { for all $v \in V$ simultaneously with probability at least $p$. }
\]
    Then $\alpha = \Omega(\gamma^{-1}\log (np))$.
\end{theorem}
\begin{proof}
    Let $d = \lceil (2\alpha+1)\gamma\rceil$  and let $G$ be any $(d+1)$-regular graph on $V$. For each vertex $v \in V$, let $G_v$ be the same as $G$, except that all $d+1$ edges incident to $v$ are removed. Observe that, in $G_v$, $v$ has coreness 0, while all other vertices $u \neq v$ have coreness at least $d$. Let $\mathcal{C}_v$ be the set of all coreness estimate vectors $\tilde{k}$ such that $\tilde{k}(v) \leq \alpha$ and $\tilde{k}(u) > \alpha$ for all $u \neq v$. Observe that $\mathcal{C}_v$ and $\mathcal{C}_u$ are disjoint for $u \neq v$. Furthermore, since $\gamma^{-1}d - \alpha = \alpha+1$,  $\mathcal{M}(G_v)$ 
 must return $\tilde{k}(v) \le \alpha$ and for all $u \ne v$, $\tilde{k}(u) \ge \alpha+1$ , i.e., an estimate in $\mathcal{C}_v$ with probability at least $p$.

    Since $G$ and $G_v$ differ by $d+1$ edges and $\mathcal{M}$ satisfies $\epsilon$-edge differential privacy, we have that $\Pr( \mathcal{M}(G) \in \mathcal{C}_v ) \geq e^{-\epsilon (d+1)} \Pr( \mathcal{M}(G_v) \in \mathcal{C}_v ) \geq e^{-\epsilon (d+1)} p$. Since $\mathcal{C}_u$ and $\mathcal{C}_v$ are disjoint for $u \neq v$, 
    $1 \geq \Pr( \bigcup_{v \in V} (\mathcal{M}(G) \in \mathcal{C}_v) ) = \sum_{v \in V} \Pr( \mathcal{M}(G) \in \mathcal{C}_v ) \geq  n e^{-\epsilon (d+1)} p$. By rearranging, $\lceil (2\alpha+1)\gamma \rceil = d \geq \ln (np)-1$. Therefore, $\alpha = \Omega(\gamma^{-1}\log (np))$.
\end{proof}

Let $V$ be a set of $2n+1$ vertices and let $x \in V$ be a fixed vertex. Consider any $\epsilon$-edge differentially private mechanism $\mathcal{M}$ that \emph{non-interactively} (i.e., in a single round) estimates the coreness of $x$ in a given graph on $V$, in the local model. We show that there is a large family of graphs on which $\mathcal{M}$ has constant probability of returning an estimate with $\Omega(\sqrt{n})$ error for $x$.

\begin{theorem}
\label{thm:non-interactive-coreness-lb}
For any constant $\epsilon > 0$, there exists a constant $0 < \eta < \frac{1}{2}$ such that the following holds. Suppose that $\mathcal{M}$ is a non-interactive $\epsilon$-edge differentially private local mechanism that estimates the coreness of a fixed vertex, $x$, in an arbitrary $n$-vertex graph such that, if $k(x)$ is the actual coreness of $x$, then the estimated coreness $\tilde{k}(x)$ of $x$ satisfies:
\[
    k(x) - \alpha \leq \tilde{k}(x) \leq k(x) + \alpha \text{ with  probability at least $\tfrac{1}{2}+\eta$. }
\]
Then there is a family of $n$-vertex graphs of size $2^{\Omega(n)}$ on which $\alpha = \Omega(\sqrt{n})$.
\end{theorem}

Our approach is to reduce to a known lower bound on the error of privately answering a linear number of random \emph{inner product queries} on a secret dataset $X \in \{0,1\}^n$. Here, two datasets $X$ and $X'$ are \emph{neighboring} if they differ in at most one coordinate, a query is specified by a vector $Q\in \mathbb{R}^n$, and the \emph{error} of a response $r$ to query $Q$ is $|r - \langle Q,X\rangle|$. \cite{DBLP:conf/icalp/EdenLRS23} were the first to use this approach to prove lower bounds in the local model.

Roughly speaking, the lower bound says that no  $\epsilon$-differentially private mechanism (with a trusted curator) can answer $O(n)$ random inner product queries in $\{-1,1\}^n$ so that, with constant probability, a large fraction of the answers have $o(\sqrt{n})$ error. (Otherwise, an ``attacker'' can use such queries to ``reconstruct'' $X$ with high accuracy, violating privacy.) In Appendix~\ref{app:lowerbound}, we formally state the lower bound and use it to prove the following modified variant, adapted to inner product queries in $\{0,1\}^n$. The idea is to convert a mechanism answering inner product queries in $\{0,1\}^n$ into a mechanism answering inner product queries in $\{-1,1\}^n$ with roughly the same error and privacy loss.

\begin{restatable}{theorem}{reconstruction}
\label{thm:reconstruction-attack}
    For any constants $\epsilon > 0$ and $\frac{1}{20} > \delta > 0$, there is a constant $0 < \eta < \frac{1}{2}$ such that no $(\epsilon,\delta)$-differentially private mechanism can answer $m = O(n)$ random inner product queries in $\{0,1\}^n$ on a secret dataset $X \in \{0,1\}^n$ such that, with probability at least $\Omega(\sqrt{\delta})$, a $(\frac{1}{2} + \eta)$-fraction of its answers have $o(\sqrt{n})$ error.
\end{restatable}

To apply Theorem~\ref{thm:reconstruction-attack}, we construct a $2\epsilon$-differentially private mechanism $\mathcal{N}$ that answers $m$ random inner product queries in $\{0,1\}^n$. For each query, $\mathcal{N}$ simulates $\mathcal{M}$ on a \emph{query graph} in which the coreness of $x$ is roughly the intended answer to the query. We show that, with constant probability, a large fraction of $\mathcal{N}$'s responses will have the same error as $\mathcal{M}$. Therefore, $\mathcal{M}$ has $\Omega(\sqrt{n})$ error with constant probability. All proofs are in \cref{sec:lb-proofs}.

\subparagraph*{Query graphs} Fix a partition $(A,B)$ of $V \setminus \{x\}$ with $|A| = |B|$ and an enumeration, $a_1,\dots,a_{n}$, of the vertices in $A$. Let $Q \in \{0,1\}^n$ be an arbitrary inner product query.  The \emph{query graph} for $Q$ on dataset $X$ is the graph $G_X(Q)$ on $V$ defined as follows.
\begin{enumerate}
    \item For all $i \in [n]$, $X_i \in \{0,1\}$ indicates whether $a_i$ is adjacent to $x$ and $Q_i \in \{0,1\}$ indicates whether $a_i$ is adjacent to (0) no vertex in $B$ or (1) every vertex in $B$.
    \item The vertices of $B$ form a clique, which $x$ is not adjacent to.
\end{enumerate}
The idea is that $Q$ is used to bound the coreness of vertices in $A$. Specfically, if $Q_i = 1$, then $a_i$ is adjacent to every vertex in $B$ and, hence, has coreness at least $n-1$. Otherwise, $a_i$ can only be adjacent to $x$, so it has coreness at most $1$. Thus, the coreness of $x$ is roughly the number of neighbors $a_i \in A$ such that $Q_i = 1$, i.e., $\langle Q, X \rangle$. (If $Q$ is the all-zero vector, then the coreness of $x$ may still be 1.)

\begin{restatable}{lemma}{corenessquerygraph}
\label{lem:coreness-query-graph}
    The coreness of $x$ in $G_X(Q)$ is either $\langle Q, X \rangle$ or $\langle Q,X \rangle +1$.
\end{restatable}

\subparagraph*{Answering random queries} 
Na\"{i}vely answering queries by simulating $\mathcal{M}$ on each of the corresponding query graphs leads to prohibitively large privacy loss. Our key observation is that the vertices whose neighborhoods in the query graphs depend on the secret dataset $X$, namely $\{x\} \cup A$, only have a few possible neighborhoods among all query graphs on $X$. This suggests that, to reduce the privacy loss, we can generate only a few messages for these vertices and \emph{reuse} them in all simulations.

\begin{observation}
\label{obs:querygraphs}
    The following holds for any secret dataset $X \in \{0,1\}^n$.
\begin{enumerate}
    \item The neighbors of $x$ are the same in all query graphs on $X$.
    \item Each vertex in $A$ only has two possible neighborhoods among all query graphs on $X$ (namely, $\{x\}$ and $B \cup \{x\}$, if $X_i = 1$, and $\emptyset$ and $B$ otherwise). 
    \item For all datasets $X$, the neighborhood of each vertex $b \in B$ in the query graph $G_X(Q)$ is a function of $Q$, i.e., $(B \setminus \{b\}) \cup \{ a_i \in A \mid Q_i = 1 \}$.
\end{enumerate}
\end{observation}

Specifically, our mechanism $\mathcal{N}$ answers $m$ random inner product queries $Q^{(1)},\dots,Q^{(m)} \in \{0,1\}^n$ on a secret dataset $X \in \{0,1\}^n$ as follows.
\begin{enumerate}
    \item Let $\mathbf{0} \in \{0,1\}^n$ and $\mathbf{1} \in \{0,1\}^n$ be the all-zeros and all-ones vectors, respectively. 
    \begin{enumerate}
        \item Run the user algorithm of $x$ (specified by $\mathcal{M}$) on its neighborhood in $G_X(\mathbf{0})$ to obtain the message $\pi_x$.
        \item For each $i \in [n]$: run the user algorithm of $a_i$ (specified by $\mathcal{M}$) on its neighborhood in $G_X(\mathbf{0})$ and $G_X(\mathbf{1})$ to obtain the messages $\pi_{a_i}(0)$ and $\pi_{a_i}(1)$, respectively.
    \end{enumerate}
    \item For each query $Q^{(j)}$, \emph{simulate} a run of $\mathcal{M}$ on $G_X(Q^{(j)})$ as follows:
    \begin{enumerate}
        \item Run the user algorithm of each vertex $b \in B$ (specified by $\mathcal{M}$) on its neighborhood in the query graph $G_X(Q^{(j)})$ to obtain the message $\pi_b^{(j)}$. 
        \item Run the server's algorithm (specified by $\mathcal{M}$) on the transcript \[\{\pi_x\} \cup \{\pi_{a_i}(Q^{(j)}_i) \mid i \in [n] \} \cup \{\pi^{(j)}_b \mid b \in B \} \] to obtain an estimate $\tilde{k}_j(x)$ of the coreness of vertex $x$ in the query graph $G_X(Q^{(j)})$. 
    \end{enumerate}
    
    \item Return $(\tilde{k}_j(x) : j \in [m])$.
\end{enumerate}

\subparagraph*{Analysis} 

Since we run the user algorithm of each vertex in $\{x\} \cup A$ on its respective neighborhood in at most two graphs and $\mathcal{M}$ is $\epsilon$-edge differentially private, the messages generated in the first step are collectively $2\epsilon$-edge differentially private. Since the pairs of graphs $G_X(\mathbf{0}), G_{X'}(\mathbf{0})$ and $G_X(\mathbf{1}), G_{X'}(\mathbf{1})$ each differ in at most one edge when datasets $X$ and $X'$ are neighboring, it follows that the first step is $2\epsilon$-differentially private. The other steps can be viewed as simply post-processing the outputs (messages) of the first step. Therefore, the entire procedure is $2\epsilon$-differentially private.

\begin{restatable}{lemma}{lbdp}
    $\mathcal{N}$ is $2\epsilon$-differentially private.
\end{restatable}

Lemma~\ref{lem:coreness-query-graph} implies that $\mathcal{N}$ produces a response to a query with additive error exceeding $\alpha+1$ only if the simulation of $\mathcal{M}$ on the corresponding query graph produces an estimate of the coreness of $x$ that has additive error exceeding $\alpha$. We note that, since the messages of vertices in $\{x\} \cup A$ are reused, the simulations are \emph{not} independent. Nonetheless, Markov's inequality can be used to bound the probability that a large number of simulations have error exceeding $\alpha$.

\begin{restatable}{lemma}{lbacc}
    Suppose that $\mathcal{M}$ has error exceeding $\alpha$ with probability at most $\beta$. Then, with probability at least $1-1/\gamma$, at least a $(1-\gamma\beta)$-fraction of the responses of $\mathcal{N}$ have error at most $\alpha+1$.
\end{restatable}

To summarize, we have shown that, if $\mathcal{M}$ estimates the coreness of a fixed vertex with error exceeding $\alpha$ with probability at most $\beta = \frac{1}{2}-\eta$, then with constant probability (say $1-1/1.001 \approx 0.001$), a large fraction ($1-1.001\beta \approx \frac{1}{2}+\eta$) of the responses of $\mathcal{M}$ have error at most $\alpha + 1$. Since $\mathcal{N}$ is $2\epsilon$-differentially private, it is $(2\epsilon,\delta)$-differentially private for any $\delta > 0$. Therefore, by the contrapositive of Theorem~\ref{thm:reconstruction-attack}, $\alpha = \Omega(\sqrt{n})$. %

\paragraph*{Funding}

\begin{wrapfigure}{r}{0.15\textwidth}
\includegraphics[width=0.13\textwidth]{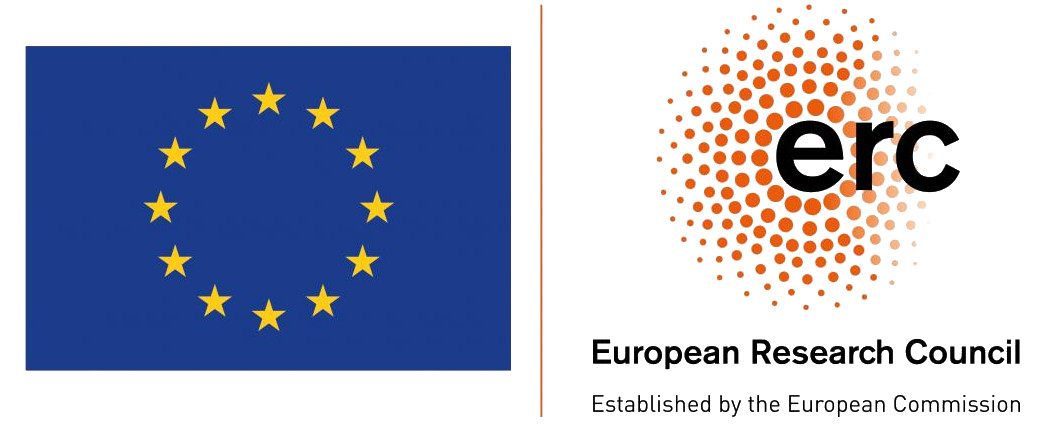}
\end{wrapfigure}

\textit{M. Henzinger and A. R. Sricharan}:
This project has received funding from the European Research Council (ERC) under the European Union's Horizon 2020 research and innovation programme (Grant agreement No. 101019564) and the Austrian Science Fund (FWF) project Z 422-N, project
I 5982-N, and project P 33775-N, with additional funding
 from the \textit{netidee SCIENCE
Stiftung}, 2020--2024. 
\bibliographystyle{gamma}
\def\bibfont{\small}
\bibliography{Ref}

\appendix

\newcommand{\M}{\mathcal{M}}
\newcommand{\A}{\mathcal{A}}
\newcommand{\HT}{\hat{\tau}}

\section{Formal definition of differential privacy in the local model}
\label{sec:prelim-extra}

We formally define our model in this section and show that private memoryless local mechanisms can be emulated using local randomizers, which shows that our results apply to the model used in prior work~\cite{DLR22kcore, DBLP:conf/icalp/EdenLRS23, dinitz2023improved}.

A \emph{local mechanism} $\mathcal{M}$ is defined by a pair of functions $f : \mathcal{T} \times \mathcal{L} \times \{0,1\}^* \to \Sigma^* \times \mathcal{L}$ and $g : \mathcal{T} \times \{0,1\}^* \to \Sigma^*$, where $\mathcal{T}$ is the set of all possible $\mathcal{M}$-transcripts\footnote{To distinguish between transcripts arising from local mechanisms and protocols using local randomizers, we will use the terminology $\M$-transcript and $\A$-transcript, respectively, in this section.}, 
$\Sigma^*$ is the set of all possible messages, $\mathcal{L}$ is the set of all possible local memory states, and $\{0,1\}^*$ is the set of all binary strings. 
An \emph{execution} of a local mechanism $\mathcal{M} = \{ f,g\}$ on an input graph $G = (V,E)$ generates a \emph{$\M$-transcript} $\hat{\tau}$ as follows. 
Initially, the local memory of each user $v$ is $\ell_v^0 := (v, N_v)$, where $N_v$ is the set of neighbors of $v$ in $G$, and the (\emph{0-round}) $\M$-transcript $\HT$ is empty. 
At the start of the $(t+1)$-th \emph{round}, each user $v$ simultaneously generates fresh random bits $\gamma_v^{t+1} \in \{0,1\}^*$ and evaluates $(\sigma_v^{t+1},\ell_v^{t+1}):=
f(\HT,\ell_v^t,\gamma_v^{t+1})$. Then the local memory of each user $v$ is set to $\ell_v^{t+1}$ and $\HT$ is updated to $\HT \odot (\sigma_v^{t+1} : v \in V)$, where
$\odot$ is the string concatenation operation. Subsequently, the server generates fresh random bits $\gamma_s^{t+1} \in \{0,1\}^*$ and $\HT$ is updated to $\HT \odot g(\HT,\gamma_s^{t+1})$. (At this point, we say $\HT$ is a \emph{$(t+1)$ $\M$-round} transcript.) If $g(\HT,\gamma_s^{t+1})$ is not the empty string, then the next round begins. Otherwise, the execution ends and the resulting $\HT$ is a \emph{complete} $\M$-transcript (on input $G$). 

We say that $\mathcal M$ is $\eps$-\emph{edge differentially private} if the output (transcript) of $\mathcal{M}$ is $\eps$-differentially private on graphs that differ by at most one edge.
We say that $\mathcal M$ is \emph{memoryless} if $f$ is the identity function on its second argument, i.e., for all $\HT,\ell,\gamma$, there exists $\sigma$ such that $f(\HT, \ell,\gamma) = (\sigma, \ell)$. In particular, in a memoryless local mechanism, the local memory of each user $v$ is always set to $(v,N_v)$.

Note that our definition implicitly assumes that the server includes each message it has received (i.e., the transcript so far) in its next ``message'' to the users (which is $\HT \odot g(\HT,\gamma_s^{t+1})$). This is why each user's ``message'' is a function of the transcript so far, even if each user is memoryless. Similarly, for each user's ``message'' to the server. Hence, since the server does not have any private data, it does not need local memory. In particular, its message can include the random string $\gamma_s^{t+1}$ it sampled.

An $\epsilon$-differentially private memoryless local mechanism $\mathcal{M} = \{f,g\}$ in our model can be modeled by a protocol $\mathcal{A}$, as defined in~\cite{DLR22kcore} and presented in Section~\ref{sec:prelim}. The main difference is that there is no server in the latter model and $\mathcal{A}$ needs to map a partial $\A$-transcript to an assignment of local randomizers to the participating users of the next round, one per user. To handle this, we hard-code the server's message in $\mathcal{M}$ that is sent at the end of the previous round into the local randomizer assigned to each user in the current round of $\mathcal{A}$. 
To simulate the random bits of the server, one fixed user $u$  sends additional random bits along with their message. This allows the users to agree on the random string generated by the server in the previous round. 

More formally, let $\mathcal{M} = \{f,g\}$ be any $\epsilon$-edge differentially private memoryless local mechanism. 
We define a protocol $\mathcal{A}$ in the local randomizer model so that each $t$-round $\A$-transcript \emph{corresponds} to a $t$-round $\M$-transcript. Indeed, the (empty) $0$-round $\A$-transcript \emph{corresponds} to the (empty) $0$-round $\M$-transcript. Now consider any $t$-round $\A$-transcript $\tau$ that corresponds to a $t$-round $\M$-transcript $\hat{\tau}$. If $\hat{\tau}$ is non-empty and the last message of the server in $\hat{\tau}$ is the empty string, then we define $\mathcal{A}(\tau) = \bot$. Otherwise, we define $\mathcal{A}(\tau) = (S_U^{t+1},S_R^{t+1},S_\epsilon^{t+1})$, where
\begin{itemize}
    \item the set $S_U^{t+1}$ of participating users  is $V$, 
    \item the local randomizer $S_R^{t+1}(v)$ assigned to user $v$ generates fresh random bits $\gamma_v \in \{0,1\}^*$ (and $\gamma_s \in \{0,1\}^*$, if $v = u$) and then outputs $S_O^{t+1}(v) = \sigma_v$ for $v \ne u$ and $S_O^{t+1}(u) = (\sigma_u,\gamma_s)$, where $(\sigma_v, (v, N_v)) = f(\hat{\tau},(v,N_v),\gamma_v)$, i.e., $\sigma_v$ is the first parameter of $f(\hat{\tau},(v,N_v),\gamma_v)$\footnote{Note the use of the $\M$-transcript $\HT$.}, and 
    \item the privacy parameter of $S_R^{t+1}(v)$ is $S_\epsilon^{t+1}(v) = \epsilon$.
\end{itemize}
In this case, the resulting $(t+1)$-round $\A$-transcript 
\[
\tau' = \tau \odot (S_U^{t+1},S_R^{t+1},S_\epsilon^{t+1},S_O^{t+1})
\]
\emph{corresponds} to the $(t+1)$-round $\M$-transcript 
\[
\hat{\tau}' = \hat{\tau} \odot (\sigma_v : v \in V) \odot g(\hat{\tau} \odot (\sigma_v : v \in V), \gamma_s),
\] 

We note that the set $\mathcal{R}$ of local randomizers used by $\mathcal{A}$ essentially consists of all functions $f$ with the first parameter hard-coded to be any value of $\mathcal{T}$, which is used to encode the corresponding $\mathcal{M}$-transcript and, in particular, the messages sent by the server to the users, as given by $g$.
Furthermore, we emphasize that, in the specification of $\mathcal{A}(\tau')$, the corresponding $\mathcal{M}$-transcript $\hat{\tau}'$ is used to determine if $\mathcal{A}(\tau') = \bot$ and, if $\mathcal{A}(\tau') \neq \bot$, to determine the output of each local randomizer. Since (the output of) each local randomizer is chosen as a function of the corresponding $\M$-transcript, it follows that $\A$ is $\epsilon$-edge differentially private.

\section{Memoryless continual counting}
\label{sec:memory_simulation}

Continual counting algorithms usually assume that one is allowed to store persistent information that can be reused between two inputs to the algorithm. We show that the binary tree mechanism can be implemented without persistent memory.

As an example, we mentioned in \cref{sec:memoryless_modif} that we use the binary tree to keep track of the degree of a vertex in a vertex-decremental graph. We wanted to keep the presence/absence of each edge private, while returning the degrees in successively smaller graphs. This was done as follows:
In each round $t$, the server sends $S_1,\dots,S_{t-1}$, the sets of vertices that were deleted in all the previous rounds. This forms a transcript $\pi_{t}$. Using this information, each vertex $v$ can determine the sequence of inputs that it previously inserted into $\mathcal{C}_v$ (namely, $|N_v|, |N_v \cap S_1|, \dots, |N_v \cap S_{t-2}|$), from which it can compute the current output $\hat \alpha_i$. The server can compute the estimated continual sum from this value as shown in the lemma below.

\begin{restatable}{lemma}{btmemoryless}
\label{lem:btmemoryless}
Given a server that sends $(i, \pi_{i, t})$ at time step $t$, where $\pi_{i, t}$ is a transcript that allows reconstructing inputs $x_{t'}$ for $t - 2^i + 1 \le t' \le t$, there is an implementation of the binary tree mechanism that is memoryless.
\end{restatable}
\begin{proof}
\cref{alg:btm} is the standard implementation of the binary tree mechanism as given in \cite{CSS10bintreemech}. Note that the only computation that needs access to the input at each time step is computing $\hat \alpha_i$. The values of $\hat \alpha_j$ for all other $j \in bin(t)$ have been calculated at previous time steps, and can be recovered from the transcript.
The server sends $i$ and a transcript that allows reconstructing the $x_{t'}$ required to produce $\hat \alpha_i$. The algorithm computes this $\hat \alpha_i$ and returns it to the server, from which the server reconstructs $\sum_{j \in bin(t)} \hat \alpha_j$.
Returning the $\hat \alpha_i$'s individually is still private, since it is simply a node in the binary tree, and the privacy proof of~\cite{CSS10bintreemech} shows that one can return all the nodes of the binary tree while being $\eps$-differentially private.
\end{proof}

\begin{minipage}[t]{0.48\textwidth}
\null
\begin{algorithm}[H]
\SetAlgoLined
\DontPrintSemicolon
\caption{Binary Tree Mechanism \cite{CSS10bintreemech}}
\label{alg:btm}
\KwInput{Time upper bound $T$, privacy parameter $\epsilon$. stream $x \in \mathbb{N}^T$}
\KwOutput{Estimate of $\sum_{t' \le t} x_{t'}$ at every $t \le T$}
$h \gets \lceil \log T \rceil$, 
$\alpha_i, \hat\alpha_i \gets 0$ for all $0 \le i \le h$\;

\For{$t = 1 \dots T$}{
$bin(t) = \{ j \mid t \pmod{2^j} \neq 0 \} $\; 
$i \gets \min \{ j \mid j \in bin(t) \}$\;
$\alpha_i \gets \sum_{t' = t - 2^i + 1}^t x_{t'}$\;
$\hat\alpha_i \gets \alpha_i + \Lap(h/\eps)$\;
}
$\alpha_j, \hat \alpha_j \gets 0$ for all $0 \le j \le i-1$\;
\Return{$\sum_{j \in bin(t)} \hat\alpha_j$}
\end{algorithm}
\end{minipage}%
\hfill
\begin{minipage}[t]{0.48\textwidth}
\null
\begin{algorithm}[H]
\SetAlgoLined
\DontPrintSemicolon
\caption{Memoryless BTM}
\label{alg:amnesiac_btm}
\KwInput{Time upper bound $T$, privacy parameter $\epsilon$. stream $x \in \mathbb{N}^T$}
\KwOutput{Estimate of $\sum_{t' \le t} x_{t'}$ at every $t \le T$}
\For{$t = 1 \dots T$}{
\textbf{receive} $i$ and transcript $\pi_{i,t}$ from the server\;
compute $x_{t'}$ for all $t - 2^i + 1 \le t' \le t$ from $\pi$\;
$\alpha_i \gets \sum_{t' = t - 2^i + 1}^t x_{t'}$\;
$\hat\alpha_i \gets \alpha_i + \Lap(h/\eps)$\;
}
\Return{$\hat\alpha_i$}
\end{algorithm}
\end{minipage}
\section{Further applications of continual counting}
\label{sec:general}

\subsection{Approximate core decomposition}
\label{sec:apx-core}

Let $\epsilon,\eta > 0$ be any positive constants. \cite{DLR22kcore} gave a memoryless local $\epsilon$-edge differentially private mechanism by building on \cite{LSY22kcore} for approximate core decomposition on $n$-vertex graphs $G = (V,E)$ with the following one-sided \emph{multiplicative} error guarantee: for every vertex $v \in V$ with actual coreness $k(v)$, the \emph{estimated} coreness $\tilde{k}(v)$ of $v$ (i.e., as reported by the mechanism) satisfies:
\[
    k(v) - \alpha \leq \tilde{k}(v) \leq (2+\eta)k(v) + \alpha \, ,
\]
where $\alpha = O(\epsilon^{-1} \log_{1+\eta} n \log^2 n)$ with probability $1-n^{-\Omega(1)}$.

Their mechanism has $T = O(\log_{1+\eta} n \log n)$ rounds, divided into $O(\log n)$ disjoint \emph{phases}, each consisting of $O(\log_{1+\eta} n)$ consecutive rounds. In each round, the non-deleted vertices send their current noisy degrees to the server. Each noisy degree is generated by adding fresh ``discrete Laplace'' noise to the corresponding actual degree. If the round occurs during phase $\phi$, then the server replies with the set of all vertices with noisy degree at most $(2+\eta)^\phi$ (in that round), which are to be deleted, and estimates the coreness of each such vertex as $(2+\eta)^{\phi-1}$. The surviving vertices update their degrees and continue to the next round.

It can be shown that, if the standard deviation of each noisy degree is $O(\epsilon^{-1}T)$, then the transcript of each round is $\epsilon T^{-1}$-edge differentially private. Therefore, by composition, the entire transcript is $\epsilon$-edge differentially private. The accuracy of the estimates follows from the following lemma, using the fact that each noisy degree differs from its corresponding actual degree by $O(\epsilon^{-1}T\log n)$ with probability $1-n^{-\Omega(1)}$. This lemma is implicitly proved in~\cite{DLR22kcore}, so we omit it.

\begin{restatable}{lemma}{apxcoresmooth}
\label{lem:apx-coreness-smooth}
    Suppose that every noisy degree differs from its corresponding actual degree by at most $\alpha$. Then, for every vertex $v \in V$ with actual coreness $k(v)$:
    \begin{itemize}
        \item the estimated coreness $\tilde{k}(v)$ of $v$ satisfies $k(v) - O(\alpha) \leq \tilde{k}(v) \leq (2+\eta) k(v) + O(\alpha)$ and
        \item every vertex in $G[U]$ has (induced) degree at least $\tilde{k}(v)/(2+\eta) - O(\alpha)$, where $U$ is the set of all vertices $u$ with estimated coreness $\tilde{k}(u) \geq \tilde{k}(v)$. 
    \end{itemize}
\end{restatable}

Similar to Section~\ref{sec:peelalg}, we observe that it is possible to obtain smaller additive error if each vertex $v$ generates a noisy initial degree using Laplace noise, and then generates its later noisy degrees using a private continual counting mechanism, $\mathcal{C}_v$, which tracks the absolute change in its degree in each round. In particular, since Observation~\ref{obs:core-sens} also holds in this setting, to ensure that the transcript is $\epsilon$-edge differentially private, it suffices to use $\mathsf{Lap}(\epsilon/4)$ noise and for each $\mathcal{C}_v$ to be $\epsilon/2$-differentially private. The privacy proof is nearly verbatim the same as Lemma~\ref{lem:peelprivacy}, so we omit it.

Each vertex makes $T = O(\log_{1+\eta} n \log n)$ insertions. Taking $\beta = (T \cdot n^{\Omega(1)})^{-1}$, the binary tree mechanism guarantees an error bound of $O(\epsinv \log n \log \log_{1+\eta} n)$ (Theorem~\ref{thm:continual-counting}). Combining this with Lemma~\ref{lem:apx-coreness-smooth} immediately yields the following.

\apxpeeling*

\subsection{Approximate densest subgraph}
\label{sec:dense}

It can be shown that if $G[U^*]$ is a densest subgraph of $G$, then the (induced) degree of every vertex in $G[U^*]$ is at least its density,  $\rho^*$. (In particular, if some vertex has degree less than $\rho^*$, then removing it yields a subgraph with higher density, which is impossible.) Hence, $G[U^*]$ is contained in the $\rho^*$-core of $G$ and $\rho^*$ is at most the maximum coreness, $k^*$, of any vertex in $G$.

\begin{lemma} [Folklore]
\label{lem:density-upperbound}
    Let $G$ be any graph. If $\rho^*$ is maximum density of any subgraph of $G$ and $k^*$ is the maximum coreness of any vertex of $G$, then $\rho^* \leq k^*$.
\end{lemma}

Let $H$ be the $k^*$-core of $G$. Since every vertex in $H$ has (induced) degree at least $k^*$, there are at least $k^*|V(H)|/2$ edges in $H$. 
Therefore, the density of $H$ is $|E(H)|/|V(H)| \geq k^*/2 \geq \rho^*/2$ and returning $H$ yields a 2-approximate solution to the densest subgraph problem on $G$.

\cite{dhulipala2023near} and \cite{DLR22kcore} leveraged the preceding observation to develop differentially private $2$-approximate and $(4+\eta)$-approximate solutions to the densest subgraph problem in the centralized and local model, respectively. 
Specificially, their mechanisms return a set of vertices $U$ and $U'$ such that the density of $G[U]$ and $G[U']$ is at least $\rho^*/2$ and $\rho^*/(4+\eta)$, respectively. Their procedures may be summarized as follows.
\begin{enumerate}
    \item[(1)] Compute an estimate $\tilde{k}(v)$ of the coreness of every vertex $v$ in $G$.
    \item[(2)] Return $\tilde{U}^* = \{ v \in V : \tilde{k}(v) = \tilde{k}^* \}$, where $\tilde{k}^* = \max \{ \tilde{k}(v) : v \in V \}$. 
\end{enumerate}

Since step (2) is simply post-processing, the entire procedure has the same differential privacy guarantees as the coreness estimation procedure employed in step (1). The next lemma gives sufficient conditions for the accuracy of the estimation to scale with the accuracy of the underlying coreness estimation procedure. Roughly speaking, \cite{dhulipala2023near} assumes that their coreness estimation procedure satisfies the second property, without proof, while~\cite{DLR22kcore} implicitly proves it.

\begin{lemma}
\label{lem:density-coreness-accuracy}
    Let $\gamma \geq 1$ be a constant. Suppose that for each vertex $v \in V$ with actual coreness $k(v)$:
    \begin{itemize}
        \item the estimated coreness $\tilde{k}(v)$ of $v$ satisfies $k(v) - \alpha \leq \tilde{k}(v) \leq \gamma k(v) + \alpha$ and
        \item every vertex in $G[U]$  has (induced) degree at least $\tilde{k}(v)/\gamma - \alpha$, where $U$ is the set of all vertices $u$ with estimated coreness $\tilde{k}(u) \geq \tilde{k}(v)$. 
    \end{itemize}
    Then the density of $G[\tilde{U}^*]$ is at least $\rho^*/2\gamma - O(\alpha)$.
\end{lemma}
\begin{proof}
    By the first assumption, $\tilde{k}^* = \max \{ \tilde{k}(v) \mid v \in V \} \geq \max \{ k(v) - \alpha \mid v \in V \} = k^* - \alpha$. By the second assumption, every vertex in $G[\tilde{U}^*]$ has induced degree at least $\tilde{k}^*/\gamma - \alpha \geq k^*/\gamma - (1+1/\gamma)\alpha$.
    It follows that $|E(G[\tilde{U}^*])| \geq |V(G[\tilde{U}^*])|(k^*/\gamma - (1+1/\gamma)\alpha)/2$ and the density of $G[\tilde{U}^*]$ is $|E(G[\tilde{U}^*])|/|V(G[\tilde{U}^*])| \geq (k^*/\gamma - (1+1/1\gamma)\alpha)/2$. By Lemma~\ref{lem:density-upperbound}, this is at least $\rho^*/2\gamma - O(\alpha)$.
\end{proof}

By employing our coreness estimation procedures from Sections~\ref{sec:peelalg} and~\ref{sec:apx-core}, which both satisfy the conditions of Lemma~\ref{lem:density-coreness-accuracy} (see Lemmas~\ref{lem:peelaccuracy} and~\ref{lem:apx-coreness-smooth}), we immediately obtain the following.

\densestsubgraph*

\section{Inner product queries}
\label{app:lowerbound}

\cite{de2012lower} proved a lower bound on the additive error of any differentially private mechanism that answers $m = O(n)$ random inner product queries in $\{-1,1\}^n$ on a secret dataset in $\{0,1\}^n$. Specifically, the lower bound says that if the error is $O(\sqrt{n})$ on a $(\frac{1}{2}+\eta)$-fraction of the responses with probability $\Omega(\sqrt{\delta})$, then the mechanism is not $(\epsilon,\delta)$-differentially private.

\begin{theorem} [Theorem 4.1 of \cite{de2012lower}]
    For any $n \in \mathbb{N}$, $\epsilon > 0$, and $1/20 > \delta > 0$, there exists positive constants $\alpha$, $\gamma$, and $\eta < 1/2$ such that any mechanism $\mathcal{M}$ that answers $m = \alpha n$ random inner product queries $Q^{(1)},\dots,Q^{(m)} \in \{-1,1\}^n$ on a secret dataset $X\in \{0,1\}^n$ satisfying
    \[
        \Pr_{\mathcal{M}, Q^{(1)},\dots,Q^{(m)}} \left[ \Pr_{i \in [m]} [ |\mathcal{M}(X)_i - \langle X, Q^{(i)} \rangle| \leq \gamma \sqrt{n} ] \geq \frac{1}{2} + \eta \right] \geq 3\sqrt{\delta} \, 
    \]
    where $\mathcal{M}(X)_i$ denotes the response of $\mathcal{M}$ on query $Q^{(i)}$, is not $(\epsilon,\delta)$-differentially private. 
\end{theorem}

To obtain the lower bound on inner product queries in $\{0,1\}^n$ that we desire, we show that any $\epsilon$-differentially private mechanism $\tilde{\mathcal{M}}$ that answers $m$ random inner product queries in $\{0,1\}^n$ can be converted into a $2\epsilon$-differentially private mechanism $\mathcal{M}$ that answers $m$ random inner product queries in $\{-1,1\}^n$. The idea is to observe that, for any $Q \in \{-1,1\}^n$, we may write $\langle Q,X \rangle = 2\langle \tilde{Q}, X \rangle - \langle \mathbf{1}, X \rangle$, where $\tilde{Q} \in \{0,1\}^n$ is such that $\tilde{Q}_i = \frac{1}{2}(\tilde{Q}_i + 1)$ and $\mathbf{1} \in \{1\}^n$ is the all-ones vector. Using this, we can simulate $\tilde{\mathcal{M}}$ on $X$, scale each response by a factor of 2, and then subtract a noisy version of $\langle \mathbf{1},X\rangle$ from each response. More precisely, to answer $m$ random inner product queries in $\{-1,1\}^n$ on $X \in \{0,1\}^n$, $\mathcal{M}$ does the following: 
\begin{enumerate}
    \item Generate $y \sim \mathsf{Lap}(1/\epsilon)$ and release $\tilde{x} = \langle \mathbf{1}, X \rangle + y$.
    \item Run $\mathcal{M}$ on $X$ to obtain responses $(r_1,r_2,\dots,r_m)$.
    \item Return $(2r_1 - \tilde{x}, 2r_2 - \tilde{x}, \dots, 2r_m - \tilde{x})$. 
\end{enumerate}

The first step is $\epsilon$-differentially private by Lemma~\ref{lem:Laplacemech}. The second step is $\epsilon$-differentially private by assumption. Therefore, by composition, the first two steps are $2\epsilon$-differentially private. The last step is simply post-processing.

Notice that since $\mathcal{M}$ answers $m$ random inner product queries $\tilde{Q}^{(1)},\dots,\tilde{Q}^{(m)} \in \{0,1\}^n$ on $X$, the vectors $Q^{(1)},\dots,Q^{(m)} \in \{-1,1\}^n$, where $Q_i^{(j)} = 2\tilde{Q}_i^{(j)} + 1$ are random. Furthermore, the error of the $j$th answer is $|(2r_j - \tilde{x}) - \langle Q^{(j)}, X\rangle| \leq |2r_j - 2\langle \tilde{Q}^{(j)}, X \rangle| + |2\langle \tilde{Q}^{(j)}, X \rangle - \langle \mathbf{1}, X \rangle - \langle Q^{(j)}, X \rangle| + |y| = 2|r_j - \langle \tilde{Q}^{(j)}, X \rangle| + |y|$. By Laplace tail bounds (Lemma~\ref{lem:laplace_tailbounds}), $|y| \leq O(\epsilon^{-1}\log n)$ with high probability. Therefore, if the error of $j$th answer of $\tilde{\mathcal{M}}$ is $o(\sqrt{n})$, then so is the error of the $j$th answer of $\mathcal{M}$, assuming $\epsilon$ is a constant. In conclusion, we obtain the following theorem.

\reconstruction* %
\section{Missing proofs from Section~\ref{sec:non-interactive}}
\label{sec:lb-proofs}

\corenessquerygraph*
\begin{proof}
Let $A' = \{ a_i \in A \mid Q_i = X_i = 1 \}$. Notice that $|A'| = \langle Q, X \rangle$. If $|A'| = 0$, then every neighbor (if any) of $x$ has degree 1. Hence, the coreness of $x$ is at most $1 = \langle Q,X \rangle +1$. 

Now suppose $|A'| \neq 0$. We will show that the coreness of $x$ equals $|A'| = \langle Q,X \rangle$. By construction, every vertex in $A'$ is adjacent to $x$ and every vertex in $B$. It follows that $A' \cup B \cup \{x\}$ induces a subgraph of $G_X(Q)$ in which every vertex has (induced) degree at least $|A'| \geq 1$. Hence, the coreness of $x$ is at least $|A'|$. 

Next consider any set of vertices $S$ containing $x$. If $S$ contains a vertex $a_i \in A \setminus A'$ that is adjacent to $x$, then $Q_i = 0$ and the degree of $a_i$ in the subgraph of $G_X(Q)$ induced by $S$ is $1 \leq |A'|$. Otherwise, every neighbor of $x$ in the subgraph of $G_X(Q)$ induced by $S$ is in $A'$ and, hence, the induced degree of $x$ is at most $|A'|$. Hence, in either case the coreness of $x$ is at most $|A'|$.
\end{proof}

\lbdp*
\begin{proof}
Let $X, X' \in \{0,1\}^n$ be two datasets that differ only at the $i$th coordinate.    Consider any valid combination of messages, $\{\pi_x\} \cup \{\pi_{a_j}(0) \mid j \in [n] \} \cup \{ \pi_{a_j}(1) \mid j \in [n] \}$, generated in the first step. Since $X$ and $X'$ differ only at the $i$th coordinate, the neighborhoods of $x$ in $G_X(\mathbf{0}),G_{X'}(\mathbf{0})$ differ in only one edge, namely $\{x,a_i\}$. 
Similarly, for the neighborhoods of $a_i$ in the pairs of graphs $G_X(\mathbf{0}),G_{X'}(\mathbf{0})$ and $G_X(\mathbf{1}),G_{X'}(\mathbf{1})$. The neighborhoods of the other vertices in $A$ are the same in each pair of graphs $G_X(\mathbf{0}), G_{X'}(\mathbf{0})$ and $G_{X}(\mathbf{1}),G_{X'}(\mathbf{1})$. It follows that the ratio of the probabilities of seeing the (partial) transcript $\{\pi_x\} \cup \{ \pi_{a_j}(0) \mid j \in [n] \}$ on the graphs $G_{X}(\mathbf{0}),G_{X'}(\mathbf{0})$ is at most $e^{\epsilon}$, since the transcript is $\epsilon$-edge differentially private. 
Similarly, for the ratio of the probabilities of seeing the (partial) transcript $\{ \pi_{a_j}(1) \mid j \in [n] \}$ on the graphs $G_X(\mathbf{1}), G_{X'}(\mathbf{1})$. Thus, the ratio of the probabilities of seeing the entire combination of messages in the first step on datasets $X$ and $X'$ is at most $e^{\epsilon} \cdot e^{\epsilon} = e^{2\epsilon}$ and the first step is $2\epsilon$-differentially private. 
The remaining steps do not require knowledge of $X$ and, hence, only serve as post-processing. Therefore, $\mathcal{N}$ is $2\epsilon$-differentially private. 
\end{proof}

\lbacc*

\begin{proof}
    For $j \in \{1,\dots,m\}$, let $X_j \in \{0,1\}$ be the indicator random variable for whether the response to the $j$th query has error exceeding $\alpha+1$, i.e., $|\tilde{k}_j(x) - \langle Q^{(j)}, X \rangle|>\alpha+1$. By Lemma~\ref{lem:coreness-query-graph}, if $k_j(x)$ is the actual coreness of $x$ in $G_X(Q^{(j)})$, then $|k_j(x) - \langle Q^{(j)},X \rangle| \leq 1$. Hence, by triangle inequality, $|\tilde{k}_j(x) - \langle Q^{(j)}, X \rangle| \leq |\tilde{k}_j(x) - k_j(x)| + |k_j(x) - \langle Q^{(j)}, X \rangle| \leq |\tilde{k}_j(x) - k_j(x)| + 1$. Since $\mathcal{M}$ has error exceeding $\alpha$ with probability at most $\beta$, it follows that $E[X_j] \leq \beta$. Hence, by linearity of expectation, $E[\sum_{j=1}^m X_j] = \sum_{j=1}^m E[X_j] \leq \beta m$ and, by Markov's inequality, $\Pr(\sum_{i=1}^m X_i \geq \gamma\beta m) \leq 1/\gamma$. 
    Therefore, with probability $1-1/\gamma$, at least $(1 - \gamma\beta) m$ responses of $\mathcal{N}$ have error at most $\alpha+1$.
\end{proof}
 
\end{document}